\documentstyle[12pt]{article}

\begin{document} 
\newcommand{\be}{\begin{equation}}
\newcommand{\ee}{\end{equation}}
\newcommand{\al}{\alpha}
\newcommand{\bt}{\beta}

\title{ Numerical analysis of renormalon technique in quantum mechanics.}

\author{ A.A.Penin${}^1$  and  A.A.Pivovarov${}^{1,2}$\\
{\it ${}^1$Institute for Nuclear Research, Russian Academy of
Sciences,}\\ {\it Moscow 117312, Russia}\\
{\it ${}^2$National Laboratory for High Energy Physics (KEK),}\\
{\it Tsukuba, Ibaraki 305, Japan}}
\date{}
\maketitle
\begin{abstract}
We discuss the ways of extracting a low energy scale of an underlying
theory using high energy scattering data. 
Within an exactly solvable model of
quantum mechanics
we analyze a
technique based on introduction of nonperturbative power corrections 
accounting for asymptotically small terms 
and 
an alternative approach exploiting a modified
running coupling constant of the model and 
nonperturbative continuation of evolution equations into an infrared 
region.
Numerical estimates show that 
the latter is more efficient
in approximating low-energy data of the model.
\end{abstract}

\vskip 1cm

\noindent PACS number(s): 11.15.Bt,  12.38.Cy
\thispagestyle{empty}
\newpage


\section{Introduction.}
Considerable and steady improvement of experimental data 
has recently caused a renewal of interest in resummation of 
perturbation theory (PT) series.
In the growing number of cases finite order theoretical 
predictions within PT have uncertainties comparable 
with experimental errors that urgently requires more accurate 
theoretical estimates. 
Nonperturbative 
(power-like) corrections to different processes within Wilson's OPE
are widely used
to match with improving of experimental data.
For cases 
that have no simple formulation in terms
of OPE
the attempts to go beyond PT are now
mostly based on using renormalons 
(for a concise up-to-date review see \cite{ZakhAkh}).
Because the expansion parameter -- 
a running coupling constant $\al_s$ -- is sufficiently large 
for moderate energies
predictions differ strongly depending on a way one chooses to handle
a strong coupling constant in the infrared region. 

The technique we further refer to as a standard one 
presently consists in
resumming bubble chaines with principal value prescription for
singularities in the Borel plane, e.g. \cite{tech}. 
Rich phenomenology can be
developed on such a base \cite{phen}
though the real sensitivity of the approach to the
infrared physics is unclear 
as well as an unambiguous disentangle of 
perturbative and nonperturbative (condensate)
contributions
\cite{MartSach}. Some other approaches use mostly the
modified running of the coupling constant \cite{Grun,Shirkov},
specific recipes for scale
setting \cite{BLM,Neub} and optimization 
of perturbation theory \cite{gru,pms}, or some modification of $\bt$
function to produce a smooth evolution at small momenta
\cite{KrasPiv}. 
Initially there is no preference between these techniques
because no exact results on the behaviour of PT in large orders or in IR
domain are known. Some general properties of the quantum field theory 
to be respected (like
analyticity) give no much help to distinguish between possibilities. Yet 
in phenomenological applications, the existing methods 
give different numerical results lying on the edge of experimental errors.
The selection of a working frame will eventually be based on
how well the particular technique fits experimental data. 
Under these
circumstances 
it seems to be 
instructive to study some simple
models \cite{MartSach,penpiv}
where different methods used in QCD could be 
quantitatively checked for their advantages 
at least for gaining some intuition useful (or misleading as well)
in more
complicated situations.

In the present paper we investigate two different
approaches within a quantum mechanical model
that mimics some general features of 
renormalons.

\section{The model.}
We consider the problem of potential scattering with 
\be 
V(r)=V_0\delta(r-r_0)
\label{pot}
\ee 
and limit ourselves to s-wave amplitudes.
Such a potential can be considered as a kind of confining (not
completely) one. 
We study a value of wave function at 
the origin (a free wave function is normalized to 
1).  The exact solution  for
scattering of the plain wave with 
momentum $k$ reads
\be
\psi(k)\equiv\psi(k,r=0)
=\left(1 + \frac{V_0}{k} e^{ikr_0}\sin(kr_0)\right)^{-1}.
\label{exact}
\ee 
where the mass of the particle in the hamiltonian is set to 1. 
To study scattering of wave packages with distributed
momentum we consider an integral of the form
\be
\Psi(\lambda)=\int_0^\infty \psi(k) W(k,\lambda){\rm d}k
\label{def}
\ee 
where 
$W(k,\lambda)$ is a normalized weight function of a package depending on 
a set of parameters $\lambda$,
$\int_0^\infty W(k,\lambda){\rm d}k=1$.
It is more convenient to deal with a function $F(\lambda)$ 
\[
\Psi(\lambda) =1 + F(\lambda) 
\]
so that $F$ vanishes if the scattering potential is switched off.
Because of oscillating factors in eq.~(\ref{exact})
integrals (\ref{def}) are not well
suitable for the PT analysis (they are ``Minkowskian'' quantities). 
For $m_{}=r_0^{-1}>|V_0|$ there are no bound states in the potential (\ref{pot})
and we can carry out the Wick rotation because
$\psi(k)$ is analytic in upper semi-plane (${\rm Im}~ k >0$)
that corresponds to the physical sheet in energy $E\sim k^2$.
In ``Euclidean'' region the exact expression for $\psi(k)$ becomes
\be 
\psi(q)=\left(1+\frac{V_0}{2q}(1-e^{-2q/m_{}})\right)^{-1}
\label{eucl}
\ee 
where $k=iq$, $q>0$.
The last formula can be obtained by PT from a Born series for the
standard Lippmann-Schwinger equation of potential
scattering since we deal with a finite range potential. 
The Born series is ordered in $V_0$ with general form of 
a n$th$ term 
$
V_0^n \Phi_n(q,m)$.
Bearing in mind the high energy analysis one can classify the behavior
of Born series according the rate with which its terms vanish at large
momenta. Then
each term of the Born series for $\psi(q)$ ($\Phi_n(q,m)$)
contains contributions vanishing as a power
and as an exponent
that is similar to the situation in QCD where 
PT contains two kinds of terms different with respect to their 
high energy behavior --  logarithmic and 
power like.
This different behavior allows one to easily separate terms
accordingly. Such a structure of PT series is in fact a justification for us 
of using this model as a toy 
analog of QCD and 
one must always remember a toy character of the model. 
The parameter $V_0$ then  
determines
the scale at which the series of power vanishing
terms becomes poorly  
convergent while the exponential terms have already died out completely.
The parameter $m_{}$ determines the scale
of exponentially supressed effects.
In contrast to QCD where in the massless limit both logatithmic and
power suppressed terms are governed by a single dimensional parameter
$\Lambda_{\rm QCD}$, in our model one can change the relative weight of 
power and exponential terms varying parameters $m$ and $V_0$
independently or rather choosing the particular ratio $V_0/m$.
Clearly at large values of this ratio exponetially small terms can
hardly be
detected at all and any modification of PT is going to be successful
because corrections are really tiny. We however consider more
interesting situation when this ratio is small enough and at some
moderate energies both types of contributions are present with
relatively essential weights. In such a situation one encounters a
dilemma whether to keep 
them still separated therefore using the classification
inspired by asymptotic considerations or 
to use more direct optimization
technique though lifting the requirement of proper asymptotic behaviour 
but
more efficient and precise at moderate energies. The numerical
investigation 
of
this problem is in fact the purpose of our paper.

Note that we do not fix the sign of the parameter $V_0$ and will study 
both attractive and repulsive interaction.

At high energies ($q\gg m_{}$) the expansion parameter 
is a running coupling constant $\alpha (q)=V_0/2q$ 
(trivial asymptotic freedom as in superrenormalizable theories)
\[
\psi(q)=\psi^{as}(\al) + \psi^{np}(m_{},V_0,q)
\]
where 
\be
\psi^{as}(\al)=\sum_{n=0}^\infty(-\al)^n 
\label{naivPT}
\ee
and $\psi^{np}(m_{},V_0,q)$ stands for 
exponentially suppressed ``nonperturbative'' terms
\be
\psi^{np}(m_{},V_0,q)={V_0\over 2q}e^{-2\frac{q}{m_{}}}+\ldots
\label{nonPT}
\ee
Within the present model we classify terms with respect to their behavior 
at large $q$:
power like vanishing -- PT, faster than any power -- non-PT.
With the standard renormalization group (RG)
terminology we have 
\begin{equation}                                
\bt(\al)=q{\partial\al\over \partial q}=-\al.
\label{b1}
\end{equation}
Resummation in eq.~(\ref{naivPT}) 
(in the spirit of RG) results in definition of 
a new (renormalization group improved) ``running'' coupling constant
\be
\al_{as}(q) = {V_0\over 2q+V_0}
\label{resumPT}
\ee
with a $\bt$ function
\be
\bt^{as}(\al_{as})=-\al_{as}(1-\al_{as}) .
\label{bas}
\ee
This $\beta$ function has an infrared fixed point that makes the running
coupling constant $\al_{as}(q)$
finite at small $q$  
in accordance with the explicit expression (\ref{resumPT}).
The use of this expansion parameter allows us to improve
the perturbation theory and to sum up all ``perturbative'' 
power terms of the series (\ref{naivPT})
\be
\psi^{as}(q)=1+\al_{as}(q).
\label{resumfn}
\ee

Now we turn to consideration of the wave 
package of a specific form
given by the following weight function
\[
W(q,Q)=Q{e^{-{Q\over q}}\over q^2}.
\]
This weight function has a bump of the width $\sqrt{3}Q$ 
located at $q\sim Q/2$
so the above  wave package can be considered as a ``probe''
of the scattering potential at the scale $\sim 2/Q$.

\section{Borel summation and renormalons.}
It is easy to see that our observable (\ref{def})
suffers from
the renormalon.
Substituting $\psi(q)$ in eq.~(\ref{def}) by its asymptotic 
expansion (\ref{resumfn}) we obtain 
\be
F^{as}(\al)= \int_0^\infty \al_{as}(q)W(q,Q){\rm d}q.
\label{df}
\ee
The quantity $F^{as}(\al)$ has a typical structure 
of QCD observable containing renormalon, {\it i.e.}
it is 
an integral of some weight function multiplied by
a running coupling constant over the interval that
includes strong coupling domain.
Note that in our model the use of the running coupling 
constant (\ref{resumPT})
in the integrand accounts for all perturbative corrections. 
Situation in 
QCD is much more complicated and the representation
of such a type is justified by the 
assumption of ``naive nonabelianization''\cite{nonab}.

After integrating
term by term in eq.~(\ref{df}) 
we get the series with factorially growing coefficients
\be
F^{as}(\al)=\sum_{n=1}^\infty (-1)^nn!\al^n, \quad \al=V_0/2Q.
\label{serf}
\ee
Properties of the series (\ref{serf}) depend crucially
on the sign of $\al$ or $V_0$.  Let us consider first repulsive 
potential $V_0>0$.
Then the alternating series~(\ref{serf}) is Borel summable
(in QCD it might correspond to an ultraviolet renormalon).
The Borel image
\be
\tilde F^{as}(u)=-{u\over 1+u} 
\label{serfb}
\ee
has a pole at $u=-1$ and is a regular function
on the positive semiaxis. 
So the Borel summation leads to the result 
($\al=V_0/2Q$)
\be
F_B(\al)= {e^{-{1/\al}}\over\al}{\rm E_1}\left({1\over\al}\right)-1 
\label{bsp}
\ee
where ${\rm E_1}(x)$ is the integral exponent~\cite{mh}
$$
{\rm E}_1(x)=\int_x^\infty{e^{-t}\over t}.  
$$
Clearly the Borel resummation procedure
in this case gives an unambiguous 
meaning to the series (\ref{serf}) and essensially improves on  
convergence of partial sums of the series (\ref{serf}). 
However numerical analysis shows that 
$F_B(Q)$ does not approximate well
the exact function $F(Q)$ for intermediate $Q\sim m_{}$ (Fig.~1). 
Indeed,  
$\al_{as}(q)$ is the best expansion parameter (exponentially accurate)
at large momenta
but $\psi^{as}(q)$ does not approximate 
well the function $\psi(q)$ at small $q$
and eq.~(\ref{bsp}) tells us nothing about the 
parameter $m_{}$ that measures exponentially suppressed 
``nonperturbative'' contributions.
Is there more efficient way of extracting information 
on the parameter $m$? 
Alternative approach is to compute the function $F(Q)$ within 
the modified perturbation theory for $\psi^{as}(q)$
that can provide a sufficient accuracy even at 
very small $q$. For this purpose one has to choose
a relevant expansion parameter. 
We write
$$
\al_{as}(q)
=\al^{(1)}_\mu(q)\left(1-{\mu-V_0\over V_0}\al^{(1)}_\mu(q)\right)^{-1}
=\al^{(1)}_\mu(q)\sum_{n=0}^\infty\left({\mu-V_0\over V_0}\right)^n
\al^{(1)}_\mu(q)^n,
$$
$$
\al^{(1)}_{\mu}(q) = {V_0\over 2q+\mu}
$$
where $\mu$ is a parameter that reflect some freedom in the choice 
of a scheme.
Now we limit ourselves to only two terms of this expansion that is reasonable 
in PT region (though the precision may be improved on
using more terms) and define new expansion parameter
\be
\al_{\mu}^{(2)}(q) =\al^{(1)}_\mu(q)\left(1+{\mu-V_0\over V_0}\al^{(1)}_\mu(q)\right)
={V_0\over 2q+\mu}\left(1+{\mu -V_0\over 2q+\mu}\right)
\label{resumPTm}
\ee
with a $\bt$ function
\[
\bt^{(2)}_\mu(\al)=-\al +O(\al^2)
\]
that is a PT transformation.
The perturbation theory series for $\psi(q)$ in $\al_{\mu}^{(2)}$
reads 
\be
\psi^{as}(q)=1+\al_{\mu}^{(2)}(q)+O(\al^2(q)).
\label{resumm}
\ee
Taking the first order term in 
eq.~(\ref{resumm}) and fixing the parameter $\mu$ 
at an optimal value 
we find the function 
$F^{\mu}(Q)=\int_0^\infty \al_{\mu}^{(2)}(q)W(q,Q){\rm d}q$ 
to be very close to the 
exact function $F(Q)$ up to very small $Q$ (Fig.~1).
Note that at very large $Q$ the function $F_B(Q)$ becomes 
closer to the exact result $F(Q)$ than $F^{\mu}(Q)$  
because $F_B(Q)$ and $F(Q)$  
have the same asymptotic expansion by construction. 
On the other hand
lifting this too strong condition
of the same asymptotic behavior at infinitely large $q$
we find a function $F^{\mu}(Q)$ 
that approximates the exact function $F(Q)$
uniformly for moderate $Q$.
Because at the optimal $\mu$ the function $\psi^{(2)}(q)$
is close to $\psi(q)$ for finite $q$ this approximation is universal
in a sense that it works well for various forms of 
scattering package.    
The optimal value $\mu^{opt}(m_{},V_0)$ can be extracted from experiment
(in our model the exact solution plays the role
of experimental data). It turns out to be very sensitive to the 
variation of $m_{}$ while the Borel resummed series is 
universal and gives the same prediction
for theories with different $m$. 
So, in this case the modification of the running 
of the coupling constant is more efficient and flexible in 
determining the low energy structure of the model.
Stress again that Borel resummed result in this case 
has no build-in mechanism to modify predictions in dependence 
on the parameter $m$ while the value of optimal parameter $\mu$
directly and rather sensitively reflects the change of $m$. 

The case of attractive potential $V(r)=-V_0\delta(r-r_0)$, $V_0 >0$ 
at first sight seems to be
completely different. 
The running coupling constant 
$$
\al_{as}(q)={V_0\over 2q-V_0}
$$
has a singularity at $q=V_0/2>0$.  
The series (\ref{serf}) now becomes
\be
F^{as}(\al)=\sum_{n=1}^\infty n!\al^n, \quad \al=V_0/2Q>0
\label{pos}
\ee
and it is not Borel summable. This situation resembles the case of 
the infrared renormalon.
The Borel image of the series (\ref{pos})
has a pole on the positive 
semiaxis
and the Borel procedure leads to an ill-defined 
representation in case of 
attractive potential.
Though this could be considered as a signal of the presence 
of nonperturbative contributions one should stress that the  
exact function (\ref{eucl})
undergoes no qualitative change in the low energy domain
(see also \cite{deRaf}).
Following the line of QCD renormalon technique
we define the result 
of Borel summation by deforming the integration contour 
in the complex $u$ plane.
The result obtained in this way depends on the specific form of
an integration contour
while an appropriate nonperturbative
part must cancel this dependence.
In our model we use the principal value (PV) prescription 
to define the sum of the series (\ref{pos})
\be
F_B(\al)
= {\rm PV}\int_0^\infty e^{-{u/\al}}{u\over 1-u}{{\rm d}u\over\al}=
{e^{-{1/\al}}\over\al}{\rm Ei}\left({1\over\al}\right)-1  
\label{bsm}
\ee
where ${\rm Ei}(x)$ is the integral exponent~\cite{mh}
$$
{\rm Ei}(x)={\rm PV}\int^x_{-\infty}{e^{t}\over t}.  
$$
$\al=V_0/2Q>0$.  
Within the renormalon technique one should search for the exact 
function $F(Q)$ in the form  
\be
F(Q)=F_B(\al)+Ce^{-{1/\al}}+\ldots  
\label{bsmc}
\ee
Here the first term has the same perturbative 
asymptotic expansion as 
the exact function $F(Q)$, a constant 
$C$ gives the leading exponentially suppressed 
correction and ellipsis stands for ``higher twist'' contributions.
The power of the exponent in~(\ref{bsmc}) is 
determined by the position of the pole of the Borel image.
To find the value of $C$ one has to use purely
nonperturbative method or extract it from experiment.
The form of first term, the value of the constant $C$, 
and high order corrections
do depend on summation prescription while the whole
sum should not by construction. 
So, in contrast to the previous 
case where the PT sum was uniquely determined by Borel prescription,
in this case there is a parameter that can 
be adjusted to fit experimental data and therefore 
to measure the value of $m$.
It happens that the parameter $C(m_{},V_0)$ 
is quite sensitive to the 
variation of $m_{}$ so it can be considered as a ``probe''
of nonperturbative effects. 

The result of approximation for 
the exact function $F(Q)$ with
eq.~(\ref{bsmc}) 
is given in Fig.~2.

The approximation diverges strongly at small $Q$.
The reason is the same as in the previous case of
the Borel summable series. Namely, 
$\al_{as}(q)$ is a poor expansion parameter at small momenta. 
Still at sufficiently large $Q$ the accuracy is reasonable and adding 
of an adjusting term (constant $C$) improves on a precision of 
the resummed series
at moderate energies. This is rather natural because an introduction of
an additional free parameter ($C$ in this case)
always allows one to get better results.

Though the running coupling $\al_{as}(q)$ becomes singular at some point
and can hardly be used for approximation of regular
function $\psi(q)$ the introduction of a new term
for Borel nonsummable series (extra degree of freedom) helps to detect
the dependence of the constant $C$ on $m$ through experimental data.
Note that this method gives no hint on how to construct
``higher twist'' contributions in this case because there is a single
singularity of the Borel image on the positive semiaxes allowing only
one additional degree of freedom.  

Stress once again however that even the singular function
$\al_{as}(q)$ still accumulates
all PT terms exactly as in the previous case. 
Though making results numerically better, in our model 
the assumption that a singularity of the Borel image determines the
index of vanishing of the leading nonperturbative term is incorrect.
Indeed eq.~(\ref{bsmc}) fails to detect the leading non-PT asymptotics
of the exact function $F(Q)$ which reads
\[
\sqrt{\al V_0\over m_{}}
\exp\left(-2 \sqrt{V_0\over \al m_{}}\right)
\]
while the Borel based guess is $exp(-1/\al)$, $\al=V_0/2Q$.
In addition, the parameterization~(\ref{bsmc})
is not universal {\it i.e.} one  gets essentially different values of $C$ 
for different  scattering packages and it is not clear how this variety
of parameters should be used to estimate the infrared parameter $m$
of the
model.   

Turning to an alternative approach and 
introducing a running coupling constant of the 
form (\ref{resumPTm}) with an appropriate $\mu$
we obtain a uniform approximation of the function $F(Q)$
practically for all $Q$ (Fig.~2). 
Because in our toy model experimental data (exact
solution)
is available for all Q with arbitrary accuracy we do not limit the
applicability of the methods in question to the asymptotic regime
only but try them at all possible Q.
Thus, the approach based on introduction of a single parameter as in the
previous
case produces more efficient fitting for experimental data in broader
range of energies. The accuracy of the approximation and the
actual value of the parameter $\mu$ (and $C$ as well)
depend on the range where data have to be fitted. The
renormalon approach cannot be used for small $Q$ by
construction of approximants while the modified running
technique is applicable till rather small energies.
The optimal value of the parameter $\mu$ depends
on the scattering packages rather weakly because this technique
essentially approximates the exact function (\ref{eucl}) pointwise. 
This dependence is also expected because it reflects
the different choice of optimal scheme for different observables.

In QCD this prescription would correspond to the use
of (probably mass dependent) RG equation for 
strong coupling constant with infrared regular 
solution~\cite{Grun,KrasPiv,Khoz}. 
We should note however that in this way 
we determine only the scale of nonperturbative effects 
while the exact form of $F(Q)$ can be found only
via real nonperturbative calculations.  

\section{Conclusion.}
The analysis of two possible ways of extracting information on low
energy domain of a quantum mechanical model
shows that approach based on optimization of the PT through introduction
of a flexible expansion parameter using 
the freedom of choice of the scheme
is more efficient for moderate energies than
direct Borel resummation technique.
In our model also the singularity of Borel image for attractive
potential
does not correspond to leading non-PT asymptotics of exact function
that is one of main reasons for using the renormalon technique in
phenomenological applications of QCD.
Though obtained in a toy model, these observations may serve as a
ground for using a modified running of the coupling constant of QCD
in the
infrared domain for phenomenological applications instead
(or in addition to) the 
standard renormalon technique.  
\vspace{0.5cm}

\noindent
{\large \bf Acknowledgments}\\[1mm]
A.A.Pivovarov thanks Prof.~Y.Okada for his interest in the work, 
comments, and discussion. A.A.Pivovarov is grateful to Prof.~Y.Okada 
and to colleagues from Theory Group
for the kind hospitality extended to him during
the stay at KEK where the final version of the paper was written.  
The work of A.A.Pivovarov is supported in part
by Russian Fund for Basic Research No. 96-01-01860.
The work of A.A.Penin is supported in part by INTAS grant N~93-1630-ext
and N~93-2492-ext (research program of 
International Fund for Fundamental Physics in
Moscow).

\vspace{0.5cm}
\noindent
{\large \bf Figure Captions}\\[1mm]
Fig.~1. Numerical results for repulsive potential $V_0=1$,
$m_{}=3$.\\
Function $F^{\mu}(Q)/F(Q)$, the optimal value $\mu =5$ (line $a$).\\
Function  $F_B(Q)/F(Q)$ (line $b$).\\
Function  $F^{(3)}(Q)/F(Q)$  where
$F^{(3)}(Q)$ is the sum of the first three terms of the
asymptotic expansion~(\ref{serf}) (line $c$).\\[1mm]
Fig.~2. Numerical results for atractive potential $V_0=1$, $m_{}=3$.  \\
Function  $F^{\mu}(Q)/F(Q)$ for the optimal value $\mu =2.7$ (line $a$).\\
Function  $(F_B(Q)+Ce^{-{1/\al}})/F(Q)$ 
for an optimal value $C=-0.06$ (line $b$).\\
Function  $F_B(Q)/F(Q) $ (line $c$).

\newpage
\vspace*{40mm}
\setlength{\unitlength}{0.240900pt}
\ifx\plotpoint\undefined\newsavebox{\plotpoint}\fi
\sbox{\plotpoint}{\rule[-0.200pt]{0.400pt}{0.400pt}}%
\begin{picture}(1500,1350)(0,0)
\font\gnuplot=cmr10 at 10pt
\gnuplot
\sbox{\plotpoint}{\rule[-0.200pt]{0.400pt}{0.400pt}}%
\put(176.0,113.0){\rule[-0.200pt]{0.400pt}{292.453pt}}
\put(176.0,234.0){\rule[-0.200pt]{4.818pt}{0.400pt}}
\put(154,234){\makebox(0,0)[r]{0.8}}
\put(1416.0,234.0){\rule[-0.200pt]{4.818pt}{0.400pt}}
\put(176.0,356.0){\rule[-0.200pt]{4.818pt}{0.400pt}}
\put(154,356){\makebox(0,0)[r]{0.9}}
\put(1416.0,356.0){\rule[-0.200pt]{4.818pt}{0.400pt}}
\put(176.0,477.0){\rule[-0.200pt]{4.818pt}{0.400pt}}
\put(154,477){\makebox(0,0)[r]{1.0}}
\put(1416.0,477.0){\rule[-0.200pt]{4.818pt}{0.400pt}}
\put(176.0,599.0){\rule[-0.200pt]{4.818pt}{0.400pt}}
\put(154,599){\makebox(0,0)[r]{1.1}}
\put(1416.0,599.0){\rule[-0.200pt]{4.818pt}{0.400pt}}
\put(176.0,720.0){\rule[-0.200pt]{4.818pt}{0.400pt}}
\put(154,720){\makebox(0,0)[r]{1.2}}
\put(1416.0,720.0){\rule[-0.200pt]{4.818pt}{0.400pt}}
\put(176.0,841.0){\rule[-0.200pt]{4.818pt}{0.400pt}}
\put(154,841){\makebox(0,0)[r]{1.3}}
\put(1416.0,841.0){\rule[-0.200pt]{4.818pt}{0.400pt}}
\put(176.0,963.0){\rule[-0.200pt]{4.818pt}{0.400pt}}
\put(154,963){\makebox(0,0)[r]{1.4}}
\put(1416.0,963.0){\rule[-0.200pt]{4.818pt}{0.400pt}}
\put(176.0,1084.0){\rule[-0.200pt]{4.818pt}{0.400pt}}
\put(154,1084){\makebox(0,0)[r]{1.5}}
\put(1416.0,1084.0){\rule[-0.200pt]{4.818pt}{0.400pt}}
\put(176.0,1206.0){\rule[-0.200pt]{4.818pt}{0.400pt}}
\put(154,1206){\makebox(0,0)[r]{1.6}}
\put(1416.0,1206.0){\rule[-0.200pt]{4.818pt}{0.400pt}}
\put(176.0,113.0){\rule[-0.200pt]{0.400pt}{4.818pt}}
\put(176,68){\makebox(0,0){0}}
\put(176.0,1307.0){\rule[-0.200pt]{0.400pt}{4.818pt}}
\put(302.0,113.0){\rule[-0.200pt]{0.400pt}{4.818pt}}
\put(302,68){\makebox(0,0){1}}
\put(302.0,1307.0){\rule[-0.200pt]{0.400pt}{4.818pt}}
\put(428.0,113.0){\rule[-0.200pt]{0.400pt}{4.818pt}}
\put(428,68){\makebox(0,0){2}}
\put(428.0,1307.0){\rule[-0.200pt]{0.400pt}{4.818pt}}
\put(554.0,113.0){\rule[-0.200pt]{0.400pt}{4.818pt}}
\put(554,68){\makebox(0,0){3}}
\put(554.0,1307.0){\rule[-0.200pt]{0.400pt}{4.818pt}}
\put(680.0,113.0){\rule[-0.200pt]{0.400pt}{4.818pt}}
\put(680,68){\makebox(0,0){4}}
\put(680.0,1307.0){\rule[-0.200pt]{0.400pt}{4.818pt}}
\put(806.0,113.0){\rule[-0.200pt]{0.400pt}{4.818pt}}
\put(806,68){\makebox(0,0){5}}
\put(806.0,1307.0){\rule[-0.200pt]{0.400pt}{4.818pt}}
\put(932.0,113.0){\rule[-0.200pt]{0.400pt}{4.818pt}}
\put(932,68){\makebox(0,0){6}}
\put(932.0,1307.0){\rule[-0.200pt]{0.400pt}{4.818pt}}
\put(1058.0,113.0){\rule[-0.200pt]{0.400pt}{4.818pt}}
\put(1058,68){\makebox(0,0){7}}
\put(1058.0,1307.0){\rule[-0.200pt]{0.400pt}{4.818pt}}
\put(1184.0,113.0){\rule[-0.200pt]{0.400pt}{4.818pt}}
\put(1184,68){\makebox(0,0){8}}
\put(1184.0,1307.0){\rule[-0.200pt]{0.400pt}{4.818pt}}
\put(1310.0,113.0){\rule[-0.200pt]{0.400pt}{4.818pt}}
\put(1310,68){\makebox(0,0){9}}
\put(1310.0,1307.0){\rule[-0.200pt]{0.400pt}{4.818pt}}
\put(1436.0,113.0){\rule[-0.200pt]{0.400pt}{4.818pt}}
\put(1436,68){\makebox(0,0){10}}
\put(1436.0,1307.0){\rule[-0.200pt]{0.400pt}{4.818pt}}
\put(176.0,113.0){\rule[-0.200pt]{303.534pt}{0.400pt}}
\put(1436.0,113.0){\rule[-0.200pt]{0.400pt}{292.453pt}}
\put(176.0,1327.0){\rule[-0.200pt]{303.534pt}{0.400pt}}
\put(806,3){\makebox(0,0){$Q/2$}}
\put(176.0,113.0){\rule[-0.200pt]{0.400pt}{292.453pt}}
\sbox{\plotpoint}{\rule[-0.400pt]{0.800pt}{0.800pt}}%
\put(1306,1262){\makebox(0,0)[r]{a}}
\put(1328.0,1262.0){\rule[-0.400pt]{15.899pt}{0.800pt}}
\put(189,842){\usebox{\plotpoint}}
\multiput(190.39,817.92)(0.536,-4.481){5}{\rule{0.129pt}{5.800pt}}
\multiput(187.34,829.96)(6.000,-29.962){2}{\rule{0.800pt}{2.900pt}}
\multiput(196.40,783.51)(0.526,-2.752){7}{\rule{0.127pt}{3.971pt}}
\multiput(193.34,791.76)(7.000,-24.757){2}{\rule{0.800pt}{1.986pt}}
\multiput(203.41,753.14)(0.509,-2.063){19}{\rule{0.123pt}{3.338pt}}
\multiput(200.34,760.07)(13.000,-44.071){2}{\rule{0.800pt}{1.669pt}}
\multiput(216.41,705.72)(0.509,-1.484){19}{\rule{0.123pt}{2.477pt}}
\multiput(213.34,710.86)(13.000,-31.859){2}{\rule{0.800pt}{1.238pt}}
\multiput(229.41,670.51)(0.509,-1.195){19}{\rule{0.123pt}{2.046pt}}
\multiput(226.34,674.75)(13.000,-25.753){2}{\rule{0.800pt}{1.023pt}}
\multiput(242.41,642.55)(0.504,-0.853){45}{\rule{0.121pt}{1.554pt}}
\multiput(239.34,645.77)(26.000,-40.775){2}{\rule{0.800pt}{0.777pt}}
\multiput(268.41,600.21)(0.504,-0.596){45}{\rule{0.121pt}{1.154pt}}
\multiput(265.34,602.61)(26.000,-28.605){2}{\rule{0.800pt}{0.577pt}}
\multiput(293.00,572.09)(0.539,-0.504){41}{\rule{1.067pt}{0.122pt}}
\multiput(293.00,572.34)(23.786,-24.000){2}{\rule{0.533pt}{0.800pt}}
\multiput(319.00,548.09)(0.728,-0.506){29}{\rule{1.356pt}{0.122pt}}
\multiput(319.00,548.34)(23.186,-18.000){2}{\rule{0.678pt}{0.800pt}}
\multiput(345.00,530.09)(0.882,-0.508){23}{\rule{1.587pt}{0.122pt}}
\multiput(345.00,530.34)(22.707,-15.000){2}{\rule{0.793pt}{0.800pt}}
\multiput(371.00,515.08)(1.123,-0.511){17}{\rule{1.933pt}{0.123pt}}
\multiput(371.00,515.34)(21.987,-12.000){2}{\rule{0.967pt}{0.800pt}}
\multiput(397.00,503.08)(1.495,-0.516){11}{\rule{2.422pt}{0.124pt}}
\multiput(397.00,503.34)(19.973,-9.000){2}{\rule{1.211pt}{0.800pt}}
\multiput(422.00,494.08)(1.797,-0.520){9}{\rule{2.800pt}{0.125pt}}
\multiput(422.00,494.34)(20.188,-8.000){2}{\rule{1.400pt}{0.800pt}}
\multiput(448.00,486.08)(2.139,-0.526){7}{\rule{3.171pt}{0.127pt}}
\multiput(448.00,486.34)(19.418,-7.000){2}{\rule{1.586pt}{0.800pt}}
\multiput(474.00,479.06)(3.951,-0.560){3}{\rule{4.360pt}{0.135pt}}
\multiput(474.00,479.34)(16.951,-5.000){2}{\rule{2.180pt}{0.800pt}}
\multiput(500.00,474.08)(3.198,-0.516){11}{\rule{4.822pt}{0.124pt}}
\multiput(500.00,474.34)(41.991,-9.000){2}{\rule{2.411pt}{0.800pt}}
\multiput(552.00,465.07)(5.597,-0.536){5}{\rule{7.133pt}{0.129pt}}
\multiput(552.00,465.34)(37.194,-6.000){2}{\rule{3.567pt}{0.800pt}}
\put(604,458.34){\rule{6.263pt}{0.800pt}}
\multiput(604.00,459.34)(13.000,-2.000){2}{\rule{3.132pt}{0.800pt}}
\put(630,456.34){\rule{6.263pt}{0.800pt}}
\multiput(630.00,457.34)(13.000,-2.000){2}{\rule{3.132pt}{0.800pt}}
\put(656,454.34){\rule{6.263pt}{0.800pt}}
\multiput(656.00,455.34)(13.000,-2.000){2}{\rule{3.132pt}{0.800pt}}
\put(682,452.84){\rule{6.263pt}{0.800pt}}
\multiput(682.00,453.34)(13.000,-1.000){2}{\rule{3.132pt}{0.800pt}}
\put(708,451.84){\rule{6.263pt}{0.800pt}}
\multiput(708.00,452.34)(13.000,-1.000){2}{\rule{3.132pt}{0.800pt}}
\put(734,450.84){\rule{6.263pt}{0.800pt}}
\multiput(734.00,451.34)(13.000,-1.000){2}{\rule{3.132pt}{0.800pt}}
\put(760,449.84){\rule{6.263pt}{0.800pt}}
\multiput(760.00,450.34)(13.000,-1.000){2}{\rule{3.132pt}{0.800pt}}
\put(812,448.84){\rule{6.263pt}{0.800pt}}
\multiput(812.00,449.34)(13.000,-1.000){2}{\rule{3.132pt}{0.800pt}}
\put(786.0,451.0){\rule[-0.400pt]{6.263pt}{0.800pt}}
\put(945,447.84){\rule{0.964pt}{0.800pt}}
\multiput(945.00,448.34)(2.000,-1.000){2}{\rule{0.482pt}{0.800pt}}
\put(838.0,450.0){\rule[-0.400pt]{25.776pt}{0.800pt}}
\put(978,447.84){\rule{0.723pt}{0.800pt}}
\multiput(978.00,447.34)(1.500,1.000){2}{\rule{0.361pt}{0.800pt}}
\put(949.0,449.0){\rule[-0.400pt]{6.986pt}{0.800pt}}
\put(1124,448.84){\rule{12.527pt}{0.800pt}}
\multiput(1124.00,448.34)(26.000,1.000){2}{\rule{6.263pt}{0.800pt}}
\put(981.0,450.0){\rule[-0.400pt]{34.449pt}{0.800pt}}
\put(1228,449.84){\rule{12.527pt}{0.800pt}}
\multiput(1228.00,449.34)(26.000,1.000){2}{\rule{6.263pt}{0.800pt}}
\put(1280,450.84){\rule{12.527pt}{0.800pt}}
\multiput(1280.00,450.34)(26.000,1.000){2}{\rule{6.263pt}{0.800pt}}
\put(1176.0,451.0){\rule[-0.400pt]{12.527pt}{0.800pt}}
\put(1384,451.84){\rule{12.527pt}{0.800pt}}
\multiput(1384.00,451.34)(26.000,1.000){2}{\rule{6.263pt}{0.800pt}}
\put(1332.0,453.0){\rule[-0.400pt]{12.527pt}{0.800pt}}
\sbox{\plotpoint}{\rule[-0.200pt]{0.400pt}{0.400pt}}%
\put(1306,1217){\makebox(0,0)[r]{b}}
\multiput(249.59,1311.97)(0.477,-4.829){7}{\rule{0.115pt}{3.620pt}}
\multiput(248.17,1319.49)(5.000,-36.487){2}{\rule{0.400pt}{1.810pt}}
\multiput(254.58,1270.96)(0.493,-3.589){23}{\rule{0.119pt}{2.900pt}}
\multiput(253.17,1276.98)(13.000,-84.981){2}{\rule{0.400pt}{1.450pt}}
\multiput(267.58,1182.84)(0.497,-2.663){49}{\rule{0.120pt}{2.208pt}}
\multiput(266.17,1187.42)(26.000,-132.418){2}{\rule{0.400pt}{1.104pt}}
\multiput(293.58,1048.26)(0.497,-1.922){49}{\rule{0.120pt}{1.623pt}}
\multiput(292.17,1051.63)(26.000,-95.631){2}{\rule{0.400pt}{0.812pt}}
\multiput(319.58,950.86)(0.497,-1.435){49}{\rule{0.120pt}{1.238pt}}
\multiput(318.17,953.43)(26.000,-71.430){2}{\rule{0.400pt}{0.619pt}}
\multiput(345.58,877.88)(0.497,-1.123){49}{\rule{0.120pt}{0.992pt}}
\multiput(344.17,879.94)(26.000,-55.940){2}{\rule{0.400pt}{0.496pt}}
\multiput(371.58,820.65)(0.497,-0.888){49}{\rule{0.120pt}{0.808pt}}
\multiput(370.17,822.32)(26.000,-44.324){2}{\rule{0.400pt}{0.404pt}}
\multiput(397.58,775.31)(0.498,-0.687){99}{\rule{0.120pt}{0.649pt}}
\multiput(396.17,776.65)(51.000,-68.653){2}{\rule{0.400pt}{0.325pt}}
\multiput(448.00,706.92)(0.498,-0.497){49}{\rule{0.500pt}{0.120pt}}
\multiput(448.00,707.17)(24.962,-26.000){2}{\rule{0.250pt}{0.400pt}}
\multiput(474.00,680.92)(0.591,-0.496){41}{\rule{0.573pt}{0.120pt}}
\multiput(474.00,681.17)(24.811,-22.000){2}{\rule{0.286pt}{0.400pt}}
\multiput(500.00,658.92)(0.723,-0.498){69}{\rule{0.678pt}{0.120pt}}
\multiput(500.00,659.17)(50.593,-36.000){2}{\rule{0.339pt}{0.400pt}}
\multiput(552.00,622.92)(0.968,-0.497){51}{\rule{0.870pt}{0.120pt}}
\multiput(552.00,623.17)(50.194,-27.000){2}{\rule{0.435pt}{0.400pt}}
\multiput(604.00,595.92)(1.249,-0.496){39}{\rule{1.090pt}{0.119pt}}
\multiput(604.00,596.17)(49.737,-21.000){2}{\rule{0.545pt}{0.400pt}}
\multiput(656.00,574.92)(1.550,-0.495){31}{\rule{1.324pt}{0.119pt}}
\multiput(656.00,575.17)(49.253,-17.000){2}{\rule{0.662pt}{0.400pt}}
\multiput(708.00,557.92)(2.043,-0.493){23}{\rule{1.700pt}{0.119pt}}
\multiput(708.00,558.17)(48.472,-13.000){2}{\rule{0.850pt}{0.400pt}}
\multiput(760.00,544.92)(2.684,-0.491){17}{\rule{2.180pt}{0.118pt}}
\multiput(760.00,545.17)(47.475,-10.000){2}{\rule{1.090pt}{0.400pt}}
\multiput(812.00,534.93)(2.999,-0.489){15}{\rule{2.411pt}{0.118pt}}
\multiput(812.00,535.17)(46.996,-9.000){2}{\rule{1.206pt}{0.400pt}}
\multiput(864.00,525.93)(3.925,-0.485){11}{\rule{3.071pt}{0.117pt}}
\multiput(864.00,526.17)(45.625,-7.000){2}{\rule{1.536pt}{0.400pt}}
\multiput(916.00,518.93)(4.649,-0.482){9}{\rule{3.567pt}{0.116pt}}
\multiput(916.00,519.17)(44.597,-6.000){2}{\rule{1.783pt}{0.400pt}}
\multiput(968.00,512.93)(5.719,-0.477){7}{\rule{4.260pt}{0.115pt}}
\multiput(968.00,513.17)(43.158,-5.000){2}{\rule{2.130pt}{0.400pt}}
\multiput(1020.00,507.94)(7.500,-0.468){5}{\rule{5.300pt}{0.113pt}}
\multiput(1020.00,508.17)(41.000,-4.000){2}{\rule{2.650pt}{0.400pt}}
\multiput(1072.00,503.94)(7.500,-0.468){5}{\rule{5.300pt}{0.113pt}}
\multiput(1072.00,504.17)(41.000,-4.000){2}{\rule{2.650pt}{0.400pt}}
\multiput(1124.00,499.95)(11.402,-0.447){3}{\rule{7.033pt}{0.108pt}}
\multiput(1124.00,500.17)(37.402,-3.000){2}{\rule{3.517pt}{0.400pt}}
\multiput(1176.00,496.95)(11.402,-0.447){3}{\rule{7.033pt}{0.108pt}}
\multiput(1176.00,497.17)(37.402,-3.000){2}{\rule{3.517pt}{0.400pt}}
\put(1228,493.17){\rule{10.500pt}{0.400pt}}
\multiput(1228.00,494.17)(30.207,-2.000){2}{\rule{5.250pt}{0.400pt}}
\put(1280,491.17){\rule{10.500pt}{0.400pt}}
\multiput(1280.00,492.17)(30.207,-2.000){2}{\rule{5.250pt}{0.400pt}}
\put(1332,489.67){\rule{12.527pt}{0.400pt}}
\multiput(1332.00,490.17)(26.000,-1.000){2}{\rule{6.263pt}{0.400pt}}
\put(1384,488.17){\rule{10.500pt}{0.400pt}}
\multiput(1384.00,489.17)(30.207,-2.000){2}{\rule{5.250pt}{0.400pt}}
\put(1328.0,1217.0){\rule[-0.200pt]{15.899pt}{0.400pt}}
\sbox{\plotpoint}{\rule[-0.500pt]{1.000pt}{1.000pt}}%
\put(1306,1172){\makebox(0,0)[r]{c}}
\multiput(1328,1172)(20.756,0.000){4}{\usebox{\plotpoint}}
\put(1394,1172){\usebox{\plotpoint}}
\multiput(313,1327)(1.851,-20.673){4}{\usebox{\plotpoint}}
\multiput(319,1260)(2.436,-20.612){5}{\usebox{\plotpoint}}
\multiput(332,1150)(3.102,-20.522){4}{\usebox{\plotpoint}}
\multiput(345,1064)(3.953,-20.376){4}{\usebox{\plotpoint}}
\multiput(358,997)(4.774,-20.199){2}{\usebox{\plotpoint}}
\multiput(371,942)(6.137,-19.827){5}{\usebox{\plotpoint}}
\multiput(397,858)(7.871,-19.205){3}{\usebox{\plotpoint}}
\multiput(422,797)(10.047,-18.162){2}{\usebox{\plotpoint}}
\multiput(448,750)(12.152,-16.826){2}{\usebox{\plotpoint}}
\multiput(474,714)(13.593,-15.685){2}{\usebox{\plotpoint}}
\multiput(500,684)(15.844,-13.407){4}{\usebox{\plotpoint}}
\multiput(552,640)(17.524,-11.121){3}{\usebox{\plotpoint}}
\multiput(604,607)(18.845,-8.698){2}{\usebox{\plotpoint}}
\multiput(656,583)(19.614,-6.789){3}{\usebox{\plotpoint}}
\multiput(708,565)(19.942,-5.753){3}{\usebox{\plotpoint}}
\multiput(760,550)(20.306,-4.296){2}{\usebox{\plotpoint}}
\multiput(812,539)(20.451,-3.540){3}{\usebox{\plotpoint}}
\multiput(864,530)(20.514,-3.156){2}{\usebox{\plotpoint}}
\multiput(916,522)(20.570,-2.769){3}{\usebox{\plotpoint}}
\multiput(968,515)(20.660,-1.987){2}{\usebox{\plotpoint}}
\multiput(1020,510)(20.694,-1.592){3}{\usebox{\plotpoint}}
\multiput(1072,506)(20.694,-1.592){2}{\usebox{\plotpoint}}
\multiput(1124,502)(20.721,-1.195){3}{\usebox{\plotpoint}}
\multiput(1176,499)(20.721,-1.195){2}{\usebox{\plotpoint}}
\multiput(1228,496)(20.740,-0.798){3}{\usebox{\plotpoint}}
\multiput(1280,494)(20.740,-0.798){2}{\usebox{\plotpoint}}
\multiput(1332,492)(20.740,-0.798){3}{\usebox{\plotpoint}}
\multiput(1384,490)(20.752,-0.399){2}{\usebox{\plotpoint}}
\put(1436,489){\usebox{\plotpoint}}
\end{picture}

\vspace{5mm}
\begin{center}
{\bf Fig. 1}
\end{center}

\newpage
\vspace*{40mm}
\setlength{\unitlength}{0.240900pt}
\ifx\plotpoint\undefined\newsavebox{\plotpoint}\fi
\sbox{\plotpoint}{\rule[-0.200pt]{0.400pt}{0.400pt}}%
\begin{picture}(1500,1350)(0,0)
\font\gnuplot=cmr10 at 10pt
\gnuplot
\sbox{\plotpoint}{\rule[-0.200pt]{0.400pt}{0.400pt}}%
\put(176.0,113.0){\rule[-0.200pt]{0.400pt}{292.453pt}}
\put(176.0,234.0){\rule[-0.200pt]{4.818pt}{0.400pt}}
\put(154,234){\makebox(0,0)[r]{0.8}}
\put(1416.0,234.0){\rule[-0.200pt]{4.818pt}{0.400pt}}
\put(176.0,356.0){\rule[-0.200pt]{4.818pt}{0.400pt}}
\put(154,356){\makebox(0,0)[r]{0.9}}
\put(1416.0,356.0){\rule[-0.200pt]{4.818pt}{0.400pt}}
\put(176.0,477.0){\rule[-0.200pt]{4.818pt}{0.400pt}}
\put(154,477){\makebox(0,0)[r]{1.0}}
\put(1416.0,477.0){\rule[-0.200pt]{4.818pt}{0.400pt}}
\put(176.0,599.0){\rule[-0.200pt]{4.818pt}{0.400pt}}
\put(154,599){\makebox(0,0)[r]{1.1}}
\put(1416.0,599.0){\rule[-0.200pt]{4.818pt}{0.400pt}}
\put(176.0,720.0){\rule[-0.200pt]{4.818pt}{0.400pt}}
\put(154,720){\makebox(0,0)[r]{1.2}}
\put(1416.0,720.0){\rule[-0.200pt]{4.818pt}{0.400pt}}
\put(176.0,841.0){\rule[-0.200pt]{4.818pt}{0.400pt}}
\put(154,841){\makebox(0,0)[r]{1.3}}
\put(1416.0,841.0){\rule[-0.200pt]{4.818pt}{0.400pt}}
\put(176.0,963.0){\rule[-0.200pt]{4.818pt}{0.400pt}}
\put(154,963){\makebox(0,0)[r]{1.4}}
\put(1416.0,963.0){\rule[-0.200pt]{4.818pt}{0.400pt}}
\put(176.0,1084.0){\rule[-0.200pt]{4.818pt}{0.400pt}}
\put(154,1084){\makebox(0,0)[r]{1.5}}
\put(1416.0,1084.0){\rule[-0.200pt]{4.818pt}{0.400pt}}
\put(176.0,1206.0){\rule[-0.200pt]{4.818pt}{0.400pt}}
\put(154,1206){\makebox(0,0)[r]{1.6}}
\put(1416.0,1206.0){\rule[-0.200pt]{4.818pt}{0.400pt}}
\put(176.0,113.0){\rule[-0.200pt]{0.400pt}{4.818pt}}
\put(176,68){\makebox(0,0){0}}
\put(176.0,1307.0){\rule[-0.200pt]{0.400pt}{4.818pt}}
\put(302.0,113.0){\rule[-0.200pt]{0.400pt}{4.818pt}}
\put(302,68){\makebox(0,0){1}}
\put(302.0,1307.0){\rule[-0.200pt]{0.400pt}{4.818pt}}
\put(428.0,113.0){\rule[-0.200pt]{0.400pt}{4.818pt}}
\put(428,68){\makebox(0,0){2}}
\put(428.0,1307.0){\rule[-0.200pt]{0.400pt}{4.818pt}}
\put(554.0,113.0){\rule[-0.200pt]{0.400pt}{4.818pt}}
\put(554,68){\makebox(0,0){3}}
\put(554.0,1307.0){\rule[-0.200pt]{0.400pt}{4.818pt}}
\put(680.0,113.0){\rule[-0.200pt]{0.400pt}{4.818pt}}
\put(680,68){\makebox(0,0){4}}
\put(680.0,1307.0){\rule[-0.200pt]{0.400pt}{4.818pt}}
\put(806.0,113.0){\rule[-0.200pt]{0.400pt}{4.818pt}}
\put(806,68){\makebox(0,0){5}}
\put(806.0,1307.0){\rule[-0.200pt]{0.400pt}{4.818pt}}
\put(932.0,113.0){\rule[-0.200pt]{0.400pt}{4.818pt}}
\put(932,68){\makebox(0,0){6}}
\put(932.0,1307.0){\rule[-0.200pt]{0.400pt}{4.818pt}}
\put(1058.0,113.0){\rule[-0.200pt]{0.400pt}{4.818pt}}
\put(1058,68){\makebox(0,0){7}}
\put(1058.0,1307.0){\rule[-0.200pt]{0.400pt}{4.818pt}}
\put(1184.0,113.0){\rule[-0.200pt]{0.400pt}{4.818pt}}
\put(1184,68){\makebox(0,0){8}}
\put(1184.0,1307.0){\rule[-0.200pt]{0.400pt}{4.818pt}}
\put(1310.0,113.0){\rule[-0.200pt]{0.400pt}{4.818pt}}
\put(1310,68){\makebox(0,0){9}}
\put(1310.0,1307.0){\rule[-0.200pt]{0.400pt}{4.818pt}}
\put(1436.0,113.0){\rule[-0.200pt]{0.400pt}{4.818pt}}
\put(1436,68){\makebox(0,0){10}}
\put(1436.0,1307.0){\rule[-0.200pt]{0.400pt}{4.818pt}}
\put(176.0,113.0){\rule[-0.200pt]{303.534pt}{0.400pt}}
\put(1436.0,113.0){\rule[-0.200pt]{0.400pt}{292.453pt}}
\put(176.0,1327.0){\rule[-0.200pt]{303.534pt}{0.400pt}}
\put(806,3){\makebox(0,0){$Q/2$}}
\put(176.0,113.0){\rule[-0.200pt]{0.400pt}{292.453pt}}
\sbox{\plotpoint}{\rule[-0.400pt]{0.800pt}{0.800pt}}%
\put(1306,1262){\makebox(0,0)[r]{a}}
\put(1328.0,1262.0){\rule[-0.400pt]{15.899pt}{0.800pt}}
\put(239,741){\usebox{\plotpoint}}
\multiput(240.41,729.93)(0.511,-1.621){17}{\rule{0.123pt}{2.667pt}}
\multiput(237.34,735.47)(12.000,-31.465){2}{\rule{0.800pt}{1.333pt}}
\multiput(252.41,695.25)(0.509,-1.236){19}{\rule{0.123pt}{2.108pt}}
\multiput(249.34,699.63)(13.000,-26.625){2}{\rule{0.800pt}{1.054pt}}
\multiput(265.41,665.79)(0.504,-0.971){43}{\rule{0.121pt}{1.736pt}}
\multiput(262.34,669.40)(25.000,-44.397){2}{\rule{0.800pt}{0.868pt}}
\multiput(290.41,619.39)(0.504,-0.724){43}{\rule{0.121pt}{1.352pt}}
\multiput(287.34,622.19)(25.000,-33.194){2}{\rule{0.800pt}{0.676pt}}
\multiput(315.41,584.58)(0.504,-0.538){43}{\rule{0.121pt}{1.064pt}}
\multiput(312.34,586.79)(25.000,-24.792){2}{\rule{0.800pt}{0.532pt}}
\multiput(339.00,560.09)(0.567,-0.505){37}{\rule{1.109pt}{0.122pt}}
\multiput(339.00,560.34)(22.698,-22.000){2}{\rule{0.555pt}{0.800pt}}
\multiput(364.00,538.09)(0.742,-0.507){27}{\rule{1.376pt}{0.122pt}}
\multiput(364.00,538.34)(22.143,-17.000){2}{\rule{0.688pt}{0.800pt}}
\multiput(389.00,521.09)(0.992,-0.504){43}{\rule{1.768pt}{0.121pt}}
\multiput(389.00,521.34)(45.330,-25.000){2}{\rule{0.884pt}{0.800pt}}
\multiput(438.00,496.08)(1.495,-0.516){11}{\rule{2.422pt}{0.124pt}}
\multiput(438.00,496.34)(19.973,-9.000){2}{\rule{1.211pt}{0.800pt}}
\multiput(463.00,487.08)(1.724,-0.520){9}{\rule{2.700pt}{0.125pt}}
\multiput(463.00,487.34)(19.396,-8.000){2}{\rule{1.350pt}{0.800pt}}
\multiput(488.00,479.08)(2.052,-0.526){7}{\rule{3.057pt}{0.127pt}}
\multiput(488.00,479.34)(18.655,-7.000){2}{\rule{1.529pt}{0.800pt}}
\multiput(513.00,472.06)(3.783,-0.560){3}{\rule{4.200pt}{0.135pt}}
\multiput(513.00,472.34)(16.283,-5.000){2}{\rule{2.100pt}{0.800pt}}
\put(538,465.34){\rule{5.200pt}{0.800pt}}
\multiput(538.00,467.34)(14.207,-4.000){2}{\rule{2.600pt}{0.800pt}}
\put(563,461.34){\rule{5.200pt}{0.800pt}}
\multiput(563.00,463.34)(14.207,-4.000){2}{\rule{2.600pt}{0.800pt}}
\put(588,457.84){\rule{6.023pt}{0.800pt}}
\multiput(588.00,459.34)(12.500,-3.000){2}{\rule{3.011pt}{0.800pt}}
\put(613,454.84){\rule{6.023pt}{0.800pt}}
\multiput(613.00,456.34)(12.500,-3.000){2}{\rule{3.011pt}{0.800pt}}
\put(638,452.34){\rule{6.023pt}{0.800pt}}
\multiput(638.00,453.34)(12.500,-2.000){2}{\rule{3.011pt}{0.800pt}}
\put(663,450.34){\rule{6.023pt}{0.800pt}}
\multiput(663.00,451.34)(12.500,-2.000){2}{\rule{3.011pt}{0.800pt}}
\put(688,448.34){\rule{6.023pt}{0.800pt}}
\multiput(688.00,449.34)(12.500,-2.000){2}{\rule{3.011pt}{0.800pt}}
\put(713,446.84){\rule{6.023pt}{0.800pt}}
\multiput(713.00,447.34)(12.500,-1.000){2}{\rule{3.011pt}{0.800pt}}
\put(738,445.84){\rule{6.023pt}{0.800pt}}
\multiput(738.00,446.34)(12.500,-1.000){2}{\rule{3.011pt}{0.800pt}}
\put(763,444.84){\rule{6.023pt}{0.800pt}}
\multiput(763.00,445.34)(12.500,-1.000){2}{\rule{3.011pt}{0.800pt}}
\put(813,443.84){\rule{6.023pt}{0.800pt}}
\multiput(813.00,444.34)(12.500,-1.000){2}{\rule{3.011pt}{0.800pt}}
\put(788.0,446.0){\rule[-0.400pt]{6.022pt}{0.800pt}}
\put(1087,443.84){\rule{12.045pt}{0.800pt}}
\multiput(1087.00,443.34)(25.000,1.000){2}{\rule{6.022pt}{0.800pt}}
\put(1137,444.84){\rule{12.045pt}{0.800pt}}
\multiput(1137.00,444.34)(25.000,1.000){2}{\rule{6.022pt}{0.800pt}}
\put(1187,445.84){\rule{11.804pt}{0.800pt}}
\multiput(1187.00,445.34)(24.500,1.000){2}{\rule{5.902pt}{0.800pt}}
\put(838.0,445.0){\rule[-0.400pt]{59.984pt}{0.800pt}}
\put(1286,446.84){\rule{12.045pt}{0.800pt}}
\multiput(1286.00,446.34)(25.000,1.000){2}{\rule{6.022pt}{0.800pt}}
\put(1336,447.84){\rule{12.045pt}{0.800pt}}
\multiput(1336.00,447.34)(25.000,1.000){2}{\rule{6.022pt}{0.800pt}}
\put(1386,448.84){\rule{12.045pt}{0.800pt}}
\multiput(1386.00,448.34)(25.000,1.000){2}{\rule{6.022pt}{0.800pt}}
\put(1236.0,448.0){\rule[-0.400pt]{12.045pt}{0.800pt}}
\sbox{\plotpoint}{\rule[-0.200pt]{0.400pt}{0.400pt}}%
\put(1306,1217){\makebox(0,0)[r]{b}}
\put(1328.0,1217.0){\rule[-0.200pt]{15.899pt}{0.400pt}}
\put(373,1273){\usebox{\plotpoint}}
\multiput(373.58,1258.96)(0.496,-4.158){45}{\rule{0.120pt}{3.383pt}}
\multiput(372.17,1265.98)(24.000,-189.978){2}{\rule{0.400pt}{1.692pt}}
\multiput(397.58,1064.18)(0.496,-3.480){43}{\rule{0.120pt}{2.848pt}}
\multiput(396.17,1070.09)(23.000,-152.089){2}{\rule{0.400pt}{1.424pt}}
\multiput(420.58,908.78)(0.496,-2.685){43}{\rule{0.120pt}{2.222pt}}
\multiput(419.17,913.39)(23.000,-117.389){2}{\rule{0.400pt}{1.111pt}}
\multiput(443.58,788.87)(0.496,-2.045){43}{\rule{0.120pt}{1.717pt}}
\multiput(442.17,792.44)(23.000,-89.435){2}{\rule{0.400pt}{0.859pt}}
\multiput(466.58,697.68)(0.496,-1.492){43}{\rule{0.120pt}{1.283pt}}
\multiput(465.17,700.34)(23.000,-65.338){2}{\rule{0.400pt}{0.641pt}}
\multiput(489.58,630.98)(0.496,-1.095){43}{\rule{0.120pt}{0.970pt}}
\multiput(488.17,632.99)(23.000,-47.988){2}{\rule{0.400pt}{0.485pt}}
\multiput(512.58,582.06)(0.496,-0.763){43}{\rule{0.120pt}{0.709pt}}
\multiput(511.17,583.53)(23.000,-33.529){2}{\rule{0.400pt}{0.354pt}}
\multiput(535.58,547.65)(0.492,-0.582){21}{\rule{0.119pt}{0.567pt}}
\multiput(534.17,548.82)(12.000,-12.824){2}{\rule{0.400pt}{0.283pt}}
\multiput(547.58,533.77)(0.492,-0.543){19}{\rule{0.118pt}{0.536pt}}
\multiput(546.17,534.89)(11.000,-10.887){2}{\rule{0.400pt}{0.268pt}}
\multiput(558.00,522.93)(0.669,-0.489){15}{\rule{0.633pt}{0.118pt}}
\multiput(558.00,523.17)(10.685,-9.000){2}{\rule{0.317pt}{0.400pt}}
\multiput(570.00,513.93)(0.692,-0.488){13}{\rule{0.650pt}{0.117pt}}
\multiput(570.00,514.17)(9.651,-8.000){2}{\rule{0.325pt}{0.400pt}}
\multiput(581.00,505.93)(0.874,-0.485){11}{\rule{0.786pt}{0.117pt}}
\multiput(581.00,506.17)(10.369,-7.000){2}{\rule{0.393pt}{0.400pt}}
\multiput(593.00,498.93)(1.155,-0.477){7}{\rule{0.980pt}{0.115pt}}
\multiput(593.00,499.17)(8.966,-5.000){2}{\rule{0.490pt}{0.400pt}}
\multiput(604.00,493.94)(1.651,-0.468){5}{\rule{1.300pt}{0.113pt}}
\multiput(604.00,494.17)(9.302,-4.000){2}{\rule{0.650pt}{0.400pt}}
\multiput(616.00,489.94)(1.505,-0.468){5}{\rule{1.200pt}{0.113pt}}
\multiput(616.00,490.17)(8.509,-4.000){2}{\rule{0.600pt}{0.400pt}}
\put(627,485.17){\rule{2.500pt}{0.400pt}}
\multiput(627.00,486.17)(6.811,-2.000){2}{\rule{1.250pt}{0.400pt}}
\put(639,483.17){\rule{2.500pt}{0.400pt}}
\multiput(639.00,484.17)(6.811,-2.000){2}{\rule{1.250pt}{0.400pt}}
\put(651,481.67){\rule{1.204pt}{0.400pt}}
\multiput(651.00,482.17)(2.500,-1.000){2}{\rule{0.602pt}{0.400pt}}
\put(656,480.67){\rule{1.445pt}{0.400pt}}
\multiput(656.00,481.17)(3.000,-1.000){2}{\rule{0.723pt}{0.400pt}}
\put(668,479.67){\rule{1.445pt}{0.400pt}}
\multiput(668.00,480.17)(3.000,-1.000){2}{\rule{0.723pt}{0.400pt}}
\put(662.0,481.0){\rule[-0.200pt]{1.445pt}{0.400pt}}
\put(720,479.67){\rule{2.650pt}{0.400pt}}
\multiput(720.00,479.17)(5.500,1.000){2}{\rule{1.325pt}{0.400pt}}
\put(674.0,480.0){\rule[-0.200pt]{11.081pt}{0.400pt}}
\multiput(743.00,481.61)(10.062,0.447){3}{\rule{6.233pt}{0.108pt}}
\multiput(743.00,480.17)(33.062,3.000){2}{\rule{3.117pt}{0.400pt}}
\multiput(789.00,484.60)(6.622,0.468){5}{\rule{4.700pt}{0.113pt}}
\multiput(789.00,483.17)(36.245,4.000){2}{\rule{2.350pt}{0.400pt}}
\multiput(835.00,488.61)(10.286,0.447){3}{\rule{6.367pt}{0.108pt}}
\multiput(835.00,487.17)(33.786,3.000){2}{\rule{3.183pt}{0.400pt}}
\multiput(882.00,491.61)(10.062,0.447){3}{\rule{6.233pt}{0.108pt}}
\multiput(882.00,490.17)(33.062,3.000){2}{\rule{3.117pt}{0.400pt}}
\put(928,493.67){\rule{5.541pt}{0.400pt}}
\multiput(928.00,493.17)(11.500,1.000){2}{\rule{2.770pt}{0.400pt}}
\put(951,494.67){\rule{5.541pt}{0.400pt}}
\multiput(951.00,494.17)(11.500,1.000){2}{\rule{2.770pt}{0.400pt}}
\put(731.0,481.0){\rule[-0.200pt]{2.891pt}{0.400pt}}
\put(997,495.67){\rule{2.891pt}{0.400pt}}
\multiput(997.00,495.17)(6.000,1.000){2}{\rule{1.445pt}{0.400pt}}
\put(974.0,496.0){\rule[-0.200pt]{5.541pt}{0.400pt}}
\put(1159,495.67){\rule{11.081pt}{0.400pt}}
\multiput(1159.00,496.17)(23.000,-1.000){2}{\rule{5.541pt}{0.400pt}}
\put(1205,494.67){\rule{11.081pt}{0.400pt}}
\multiput(1205.00,495.17)(23.000,-1.000){2}{\rule{5.541pt}{0.400pt}}
\put(1251,493.67){\rule{11.081pt}{0.400pt}}
\multiput(1251.00,494.17)(23.000,-1.000){2}{\rule{5.541pt}{0.400pt}}
\put(1297,492.67){\rule{11.322pt}{0.400pt}}
\multiput(1297.00,493.17)(23.500,-1.000){2}{\rule{5.661pt}{0.400pt}}
\put(1344,491.67){\rule{11.081pt}{0.400pt}}
\multiput(1344.00,492.17)(23.000,-1.000){2}{\rule{5.541pt}{0.400pt}}
\put(1390,490.67){\rule{11.081pt}{0.400pt}}
\multiput(1390.00,491.17)(23.000,-1.000){2}{\rule{5.541pt}{0.400pt}}
\put(1009.0,497.0){\rule[-0.200pt]{36.135pt}{0.400pt}}
\sbox{\plotpoint}{\rule[-0.500pt]{1.000pt}{1.000pt}}%
\put(1306,1172){\makebox(0,0)[r]{c}}
\multiput(1328,1172)(20.756,0.000){4}{\usebox{\plotpoint}}
\put(1394,1172){\usebox{\plotpoint}}
\multiput(411,1327)(3.237,-20.502){3}{\usebox{\plotpoint}}
\multiput(420,1270)(3.987,-20.369){12}{\usebox{\plotpoint}}
\multiput(466,1035)(5.543,-20.002){4}{\usebox{\plotpoint}}
\multiput(489,952)(6.739,-19.631){3}{\usebox{\plotpoint}}
\multiput(512,885)(8.133,-19.096){3}{\usebox{\plotpoint}}
\multiput(535,831)(9.615,-18.394){3}{\usebox{\plotpoint}}
\multiput(558,787)(11.748,-17.111){4}{\usebox{\plotpoint}}
\multiput(604,720)(14.521,-14.830){3}{\usebox{\plotpoint}}
\multiput(651,672)(16.518,-12.568){3}{\usebox{\plotpoint}}
\multiput(697,637)(17.729,-10.792){2}{\usebox{\plotpoint}}
\multiput(743,609)(18.881,-8.620){3}{\usebox{\plotpoint}}
\multiput(789,588)(19.469,-7.195){2}{\usebox{\plotpoint}}
\multiput(835,571)(19.892,-5.925){2}{\usebox{\plotpoint}}
\multiput(882,557)(20.083,-5.239){3}{\usebox{\plotpoint}}
\multiput(928,545)(20.369,-3.985){2}{\usebox{\plotpoint}}
\multiput(974,536)(20.449,-3.556){2}{\usebox{\plotpoint}}
\multiput(1020,528)(20.519,-3.122){2}{\usebox{\plotpoint}}
\multiput(1066,521)(20.639,-2.196){3}{\usebox{\plotpoint}}
\multiput(1113,516)(20.634,-2.243){2}{\usebox{\plotpoint}}
\multiput(1159,511)(20.677,-1.798){2}{\usebox{\plotpoint}}
\multiput(1205,507)(20.677,-1.798){2}{\usebox{\plotpoint}}
\multiput(1251,503)(20.712,-1.351){3}{\usebox{\plotpoint}}
\multiput(1297,500)(20.737,-0.882){2}{\usebox{\plotpoint}}
\multiput(1344,498)(20.712,-1.351){2}{\usebox{\plotpoint}}
\multiput(1390,495)(20.736,-0.902){2}{\usebox{\plotpoint}}
\put(1436,493){\usebox{\plotpoint}}
\end{picture}

\vspace{5mm}
\begin{center}
{\bf Fig. 2}
\end{center}

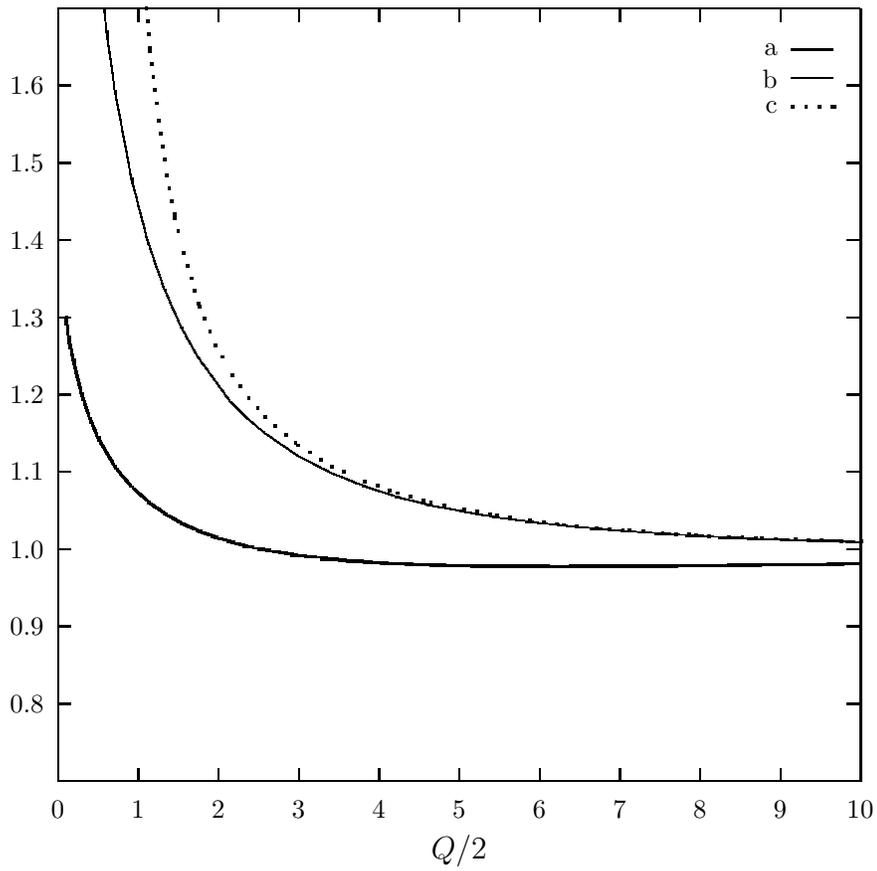
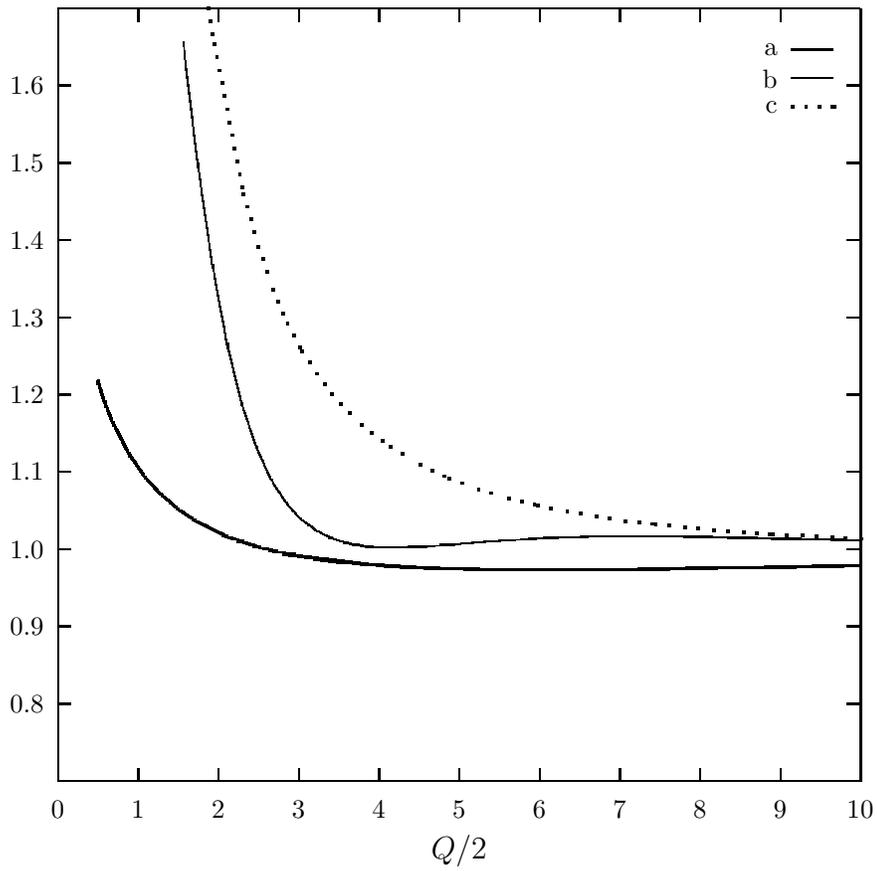
\end{document}